\documentclass[aps,prd,floatfix,twocolumn,reprint,amsmath,amssymb,superscriptaddress]{revtex4-1}
\usepackage{amsfonts}
\usepackage{mathrsfs}
\usepackage{amsmath}
\usepackage{color}
\usepackage{natbib}
\usepackage{graphicx}
\usepackage{bm}
\usepackage{amssymb}
\usepackage{mathptmx}
\usepackage{stmaryrd}
\usepackage{xspace}
\usepackage{epstopdf}
\usepackage{dcolumn}
\usepackage{longtable}
\usepackage{multirow}
\usepackage[colorlinks=true, letterpaper=true, pdfstartview=FitV, linkcolor=blue, citecolor=blue, urlcolor=blue]{hyperref}

\makeatletter

\newcommand{\Rmnum}[1]{\expandafter\@slowromancap\romannumeral #1@}

\makeatother

\begin{document}

\title{Spin-Orbit-Induced Topological Flat Bands in Line and Split Graphs of Bipartite Lattices}

\author{Da-Shuai Ma}
\affiliation{Department of Physics, Princeton University, Princeton, New Jersey, 08540, USA}
\affiliation{Beijing Key Laboratory of Nanophotonics and Ultrafine Optoelectronic Systems, School of Physics, Beijing Institute of Technology, Beijing 100081, China}

\author{Yuanfeng Xu}
\affiliation{Max Planck Institute of Microstructure Physics, 06120 Halle, Germany}

\author{Christie S. Chiu}
\affiliation{Department of Electrical Engineering, Princeton University, Princeton, New Jersey, 08540, USA}
\affiliation{Princeton Center for Complex Materials, Princeton University, Princeton, New Jersey, 08540, USA}

\author{Nicolas Regnault}
\affiliation{Laboratoire de Physique de l'Ecole normale sup\'{e}rieure, ENS, Universit\'{e} PSL, CNRS, Sorbonne Universit\'{e}, Universit\'{e} Paris-Diderot, Sorbonne Paris Cit\'{e}, Paris, France}
\affiliation{Department of Physics, Princeton University, Princeton, New Jersey, 08540, USA}

\author{Andrew A. Houck}
\affiliation{Department of Electrical Engineering, Princeton University, Princeton, New Jersey, 08540, USA}

\author{Zhida Song}
\affiliation{Department of Physics, Princeton University, Princeton, New Jersey, 08540, USA}

\author{B. Andrei Bernevig}
\email{bernevig@princeton.edu}
\affiliation{Department of Physics, Princeton University, Princeton, New Jersey, 08540, USA}

\begin{abstract}
 Topological flat bands, such as the band in twisted bilayer graphene, are becoming a promising platform to study topics such as correlation physics, superconductivity, and transport. In this work, we introduce a generic approach to construct two-dimensional (2D) topological quasi-flat bands from line graphs and split graphs of bipartite lattices. A line graph or split graph of a bipartite lattice exhibits a set of flat bands and a set of dispersive bands. The flat band connects to the dispersive bands through a degenerate state at some momentum. We find that, with spin-orbit coupling (SOC), the flat band becomes quasi-flat and gapped from the dispersive bands. By studying a series of specific line graphs and split graphs of bipartite lattices, we find that (i) if the flat band (without SOC) has inversion or $C_2$ symmetry and is non-degenerate, then the resulting quasi-flat band must be topologically nontrivial, and (ii) if the flat band (without SOC) is degenerate, then there exists an SOC potential such that the resulting quasi-flat band is topologically nontrivial. This generic mechanism serves as a paradigm for finding topological quasi-flat bands in 2D crystalline materials and meta-materials.
\end{abstract}

\date{\today}
\maketitle

\section{Introduction}

New developments in the field of many-body condensed matter physics, such as twisted bilayer graphene (TBLG) \cite{dos2007graphene,PhysRevB.82.121407,bistritzer2011moire,cao2018correlated,cao2018unconventional,lu2019superconductors,xie2019spectroscopic,sharpe2019emergent} have underlined the importance of flat bands in realizing superconductivity and magnetism.
In TBLG, a series of almost flat bands show a remarkable series of superconducting and magnetic states \cite{cao2018unconventional,PhysRevLett2018.121.087001,PhysRevB.98.085435,PhysRevB.98.085436,PhysRevX2018.8.031089,PhysRevX2018.8.041041,PhysRevLett2018.121.257001,laksono2018singlet,PhysRevLett.121.217001,PhysRevLett.121.257001,PhysRevB.98.195101,PhysRevB.98.220504,PhysRevB.98.241407,guinea2018electrostatic,PhysRevB.99.121407,PhysRevLett.122.026801,PhysRevLett.122.257002,PhysRevLett.122.026801,PhysRevLett.122.246402,yankowitz2019tuning,lu2019superconductors,sharpe2019emergent,huang2019antiferromagnetically,wu2018emergent}. It is however known \cite{BasovChubukovNaturePhysics7(4)272-276} that flat bands in Ginzburg Landau theory result in a vanishing superfluid weight, and hence no superconductivity. This is due to the fact that most flat bands are localized, the flatness usually resulting from atomic-like orbitals. It was recently argued that topology can save a flat band's superfluid weight: Chern bands support a lower bound on the superfluid density \cite{peotta2015superfluidity}, while a more exotic type of topology, present in TBLG \cite{PhysRevLett.123.036401,PhysRevX.9.021013,PhysRevB.99.195455} that exhibits zero Chern number, can also place a lower bound on the superfluid weight \cite{PhysRevLett.124.167002,PhysRevB.101.060505,peri2020fragile}. Heuristically, topological bands contain extended states, which participate in the superconductivity \cite{ryu2010topological,PhysRevB.83.220503,RevModPhys.83.1057,bernevig2013topological,sato2017topological,PhysRevLett.109.266402,PhysRevLett.117.047001,PhysRevLett.125.017001}. As such, it is important to build flat bands with topological properties.

In this work, we present one generic method of building topological flat bands in crystals with spin-orbit coupling (SOC).
A large number of these so-obtained topological flat bands are strong topological and exhibit the quantum spin Hall (QSH) effect, and the others are spinful fragile topological bands.
Fragile topological flat bands also have been found in SOC-free systems \cite{christie}, based on line graphs of non-bipartite lattices.
It is well-known that both line- and split graph lattices exhibit flat bands in their spectra \cite{mielke1991ferromagnetic,mielke1991ferromagnetism,mielke1993ferromagnetism}.
These bands are generally thought to be spanned by localized states \cite{mielke1991ferromagnetic} or contain a delocalized state due to a metallic band touch \cite{PhysRevB.78.125104}.
However, we find that if certain symmetries are maintained, the states in the flat band cannot be localized and are topological.
By adding spin-orbit coupling to line- and split graphs of bipartite lattices, we gap the previously flat but gapless exact flat band into an quasi-flat band that is topological.
This provides us with a generic way to obtain flat bands with non-trivial topology.

The remainder of this paper is organized as follows.
In Sec.\,\ref{Secii}, we discuss in detail the topological properties of the flat band in the Kagome lattice with SOC as an example.
In Sec.\,\ref{Seciii}, we identify the topological flat bands of line graphs of split graph lattices.
We conclude with a discussion in Sec.\,\ref{Secv}.
Additional specific examples of line graphs and split graphs of bipartite lattices are shown in detail in App.\,\ref{AppC}-\ref{AppE}.

\section{Non-trivial Flat Bands in the Kagome Lattice}\label{Secii}

A graph ($X$) is bipartite if all of the vertices in the graph can be divided into two sets, $U$ and $V$, such that the edges in $X$ always connect a vertex in $U$ to a vertex in $V$.
The honeycomb lattice is a well-known bipartite lattice with the two sets of vertices being the two sublattices.
A line graph $L(X)$ of a graph $X$, which we will refer to as the root graph, is constructed by replacing each edge $e_{X,i}$ in $X$ with a vertex $v_{L(X),i}$, and connecting vertex pairs $v_{L(X),i}$ and $v_{L(X),j}$ for adjacent $e_{X,i}$ and $e_{X,j}$.
As schematically shown in Fig.\,\ref{fig1}(a), the line graph of the honeycomb lattice is the Kagome lattice.
A split graph $S(X)$ is constructed from a root graph $X$ by placing an additional vertex on each edge $e_{X,i}$, as exemplified in Figs.\,\ref{fig3}(a) and \,\ref{fig3}(b).

Additional details about the bipartite lattice, line graph, and split graph are discussed in App.\,\ref{AppA}.
Here we only list some basic properties of these graphs \cite{cvetkovic2004spectral,kollar2019line}.
We only consider 2D root graphs $X$ whose edges do not cross each other.
For a bipartite lattice $X$ with $m$ polygon faces per unit cell, we have the following properties:

\noindent (i). $X$ is a bipartite lattice if and only if all the polygons in $X$ are even-sided.

\noindent (ii). The band structure of $L(X)$ consists of a set of dispersive bands plus an additional set of flat bands at $E=-2t$, where $t$ is the hopping strength between two adjacent vertices and will be set as $-1$ in this paper. The degeneracy of the flat bands of $L(X)$ is $D=m$.

\noindent (iii). Its split graph $S(X)$ is always bipartite.

\noindent (iv). There are a set of flat bands with degeneracy $D=m$ at $E=0$ in the energy spectrum of $L(S(X))$.

\noindent (v). The flat bands of both $L(X)$, $S(X)$, and $L(S(X))$ always touch the dispersive bands through a more highly degenerate state at some high-symmetry momenta.

\subsection{Kagome Lattice without SOC}
\begin{figure}[h]
\includegraphics[width=8.6cm]{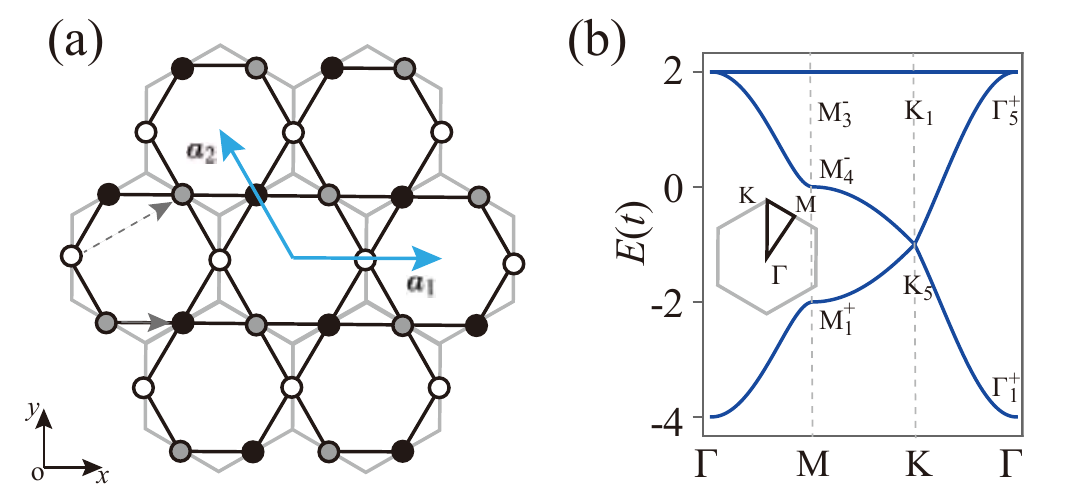}
\caption{(a) Schematic of the Kagome lattice (black). One can obtain the lattice by applying the line graph operation on the honeycomb lattice (light gray). The light blue arrows indicate the lattice vectors $\boldsymbol{a_{1}}$ and $\boldsymbol{a_{2}}$. The white, black, and gray dots represent sites in the A, B, and C sublattices, respectively. Nearest neighbor (next nearest neighbor) SOC is introduced via hopping along the gray (dashed gray) arrow direction with amplitude $i\lambda_{NN}$ ($i\lambda_{NNN}$). (b) The band structure for the Kagome lattice without SOC, with the Brillouin zone in the inset. The irreducible representations (irreps) of each bands at the high symmetry points are shown. The superscript $+/-$ denotes the parity.
\label{fig1}}
\end{figure}

The Kagome lattice is the line graph of the honeycomb lattice, which is a bipartite lattice, as depicted in  Fig.\,\ref{fig1}(a).
There are three sublattices A, B and C in the Kagome lattice. The three atoms of these sublattices are located at $\left(\frac{1}{2},0\right)$,
$\left(0,\frac{1}{2}\right)$, and $\left(\frac{1}{2},\frac{1}{2}\right)$ within each unit cell, respectively.
Here the coordinates are in units of the lattice vectors $\boldsymbol{a_{1}}=a\left( 1,0\right) $ and $\boldsymbol{a_{2}}=a\left( -1,\sqrt{3}\right) /2$ with $a$ being the lattice parameter.

The tight-binding model of the spinless lattice reads,
\begin{eqnarray}
H_{0}	=	t\sum_{\left\langle i,j\right\rangle }\left(c_{i}^{\dagger}c_{j}+\mathrm{h.c.}\right),
\end{eqnarray}
where $t=-1$ is the nearest neighbor hopping, $\left\langle i,j\right\rangle$ denotes nearest-neighbor pairs, and $c_{i}^{\dagger}$ is the creation operator of an electron on lattice site $i$. One can diagonalize the model in momentum space,
and the explicit form of the model Hamiltonian is then \:
\begin{eqnarray}
H_{0}\left(\boldsymbol{k}\right) & = & -2\left(\begin{array}{ccc}
0 & \mathrm{cos}~\boldsymbol{k}_{3} & \mathrm{cos}~\boldsymbol{k}_{2}\\
\mathrm{cos}~\boldsymbol{k}_{3} & 0 & \mathrm{cos}~\boldsymbol{k}_{1}\\
\mathrm{cos}~\boldsymbol{k}_{2} & \mathrm{cos}~\boldsymbol{k}_{1} & 0
\end{array}\right), \label{hami}
\end{eqnarray}
with $\boldsymbol{k}_{i}=\boldsymbol{k}\cdot\boldsymbol{a}_{i}/2$ and $\boldsymbol{a}_{3}=-\boldsymbol{a}_{1}-\boldsymbol{a}_{2}$.

The band structure of the Kagome lattice is shown in Fig.\,\ref{fig1}(b).
The band structure consists of a single flat band and two dispersive bands. The flat band touches one of the dispersive bands at the $\mathrm{\Gamma}$ point.

The development of topological quantum chemistry (TQC) \cite{bradlyn2017topological,elcoro2017double,PhysRevE.96.023310,PhysRevB.97.035139,PhysRevLett.121.126402} enables an efficient way to diagnose the topological phases from the symmetry-data vector, defined in App.\,\ref{AppB}, of Bloch states at high-symmetry momenta.  From TQC, the symmetry-data vector of any set of bands that cannot be decomposed into a linear combination of elementary band representations (EBRs), which are topologically equivalent with atomic orbitals in terms of the symmetry-data vector, is topological \cite{bradlyn2017topological}.
In the present work, we analyze the topological properties of the models in the language of TQC.

The model in Eq. \ref{hami} consists of spinless $s$ orbitals centered at the Wyckoff position $3f$ of the space group $P6/mmm$ (space group \#191).
The band representation (BR) of the full set of bands in Fig.\,\ref{fig1}(b) is  $\{\mathrm{\Gamma}_{1}^{+}\oplus\mathrm{\Gamma}_{5}^{+},\mathrm{K}_{1}\oplus\mathrm{K}_{5},\mathrm{M}_{1}^{+}\oplus\mathrm{M}_{3}^{-}\oplus\mathrm{M}_{4}^{-}\}$.
The character table of each irreducible representation (irrep) forming this BR is given in App.\,\ref{AppB}.
From TQC, this BR is a single EBR $\left(A_{g}\right)_{3f}\uparrow G$, which is induced from the $A_{g}$ orbital at the Wyckoff position $3f$ of the space group $P6/mmm$.
As shown in Fig.\,\ref{fig1}(b), the irrep of the flat band at the M point is $\mathrm{M}_{3}^{-}$ with parity of $-1$. There is additionally a band touching between the flat band and a dispersive band at $\Gamma$ with a two-dimensional (2D) irrep $\mathrm{\Gamma}_5^+$, of which the parity is $+2$.
In the presence of SOC, the 2D spinless irrep $\mathrm{\Gamma}_5^+$ splits into two 2D spinful irreps with parity of $+2$, respectively.
For spinful systems with inversion symmetry, the $\mathbb{Z}_{2}$ topological index $\nu$ is defined as $\left(-1\right)^{\nu}=\prod_{2n,j}P_{2n,j}$, where $P_{2n,j}$ is the parity of the $2n$-th valence band at the $j$-th time-reversal invariant momentum (TRIM) \cite{PhysRevB.76.045302}.
As any perturbative SOC does not change the parities of the bands at both $\Gamma$ and $\mathrm{M}$ points, once the band touching at the $\Gamma$ point is gapped by the symmetry-preserving SOC, one will always obtain one topologically non-trivial quasi-flat band with $\nu=1$, resulting in a QSH insulator.

\subsection{Kagome Lattice with SOC}

To identify the topology of the Kagome lattice with SOC, we expand the basis of the model in Eq. \ref{hami} to $\left\{ \begin{array}{c}
A,B,C\end{array}\right\} \varotimes\left\{ \begin{array}{c}
\uparrow,\downarrow\end{array}\right\} $ to include the spin degree of freedom. As schematically shown in  Fig.\,\ref{fig1}(a), we take both the nearest neighbor (NN) and the next nearest neighbor (NNN) SOC with respective amplitudes $i\lambda_{NN}$ and $i\lambda_{NNN}$ into account. Then, the spinful model of the Kagome lattice reads \cite{PhysRevB.86.195129},
\begin{eqnarray}
H\left(\boldsymbol{k}\right) & = & H_{0}\left(\boldsymbol{k}\right)\otimes\sigma_{0}+\left[H_{NN}\left(\boldsymbol{k}\right)+H_{NNN}\left(\boldsymbol{k}\right) \right]\otimes\sigma_{z},\label{eq5}
\end{eqnarray}
with,
\begin{eqnarray}
H_{NN}\left(\boldsymbol{k}\right)	&=&	i2\lambda_{NN}\left(\begin{array}{ccc}
0 & \mathrm{-cos}\boldsymbol{k}_{3} & \mathrm{cos}\boldsymbol{k}_{2}\\
\mathrm{-cos}\boldsymbol{k}_{3} & 0 & -\mathrm{cos}\boldsymbol{k}_{1}\\
\mathrm{cos}\boldsymbol{k}_{2} & -\mathrm{cos}\boldsymbol{k}_{1} & 0
\end{array}\right),
\end{eqnarray}
and
\begin{eqnarray}
H_{NNN}\left(\boldsymbol{k}\right)&	=	&i2\lambda_{NNN}\left(\begin{array}{ccc}
0 & \mathrm{-cos}\boldsymbol{k}_{1}^{'} & \mathrm{cos}\boldsymbol{k}_{2}^{'}\\
\mathrm{-cos}\boldsymbol{k}_{1}^{'} & 0 & -\mathrm{cos}\boldsymbol{k}_{3}^{'}\\
\mathrm{cos}\boldsymbol{k}_{2}^{'} & -\mathrm{cos}\boldsymbol{k}_{3}^{'} & 0
\end{array}\right),
\end{eqnarray} where $\boldsymbol{k}_{i}^{'}=\boldsymbol{k}_{j}-\boldsymbol{k}_{k}~\left(i\neq j\neq k\right)$.

\begin{figure}[t]
\includegraphics[width=8.6cm]{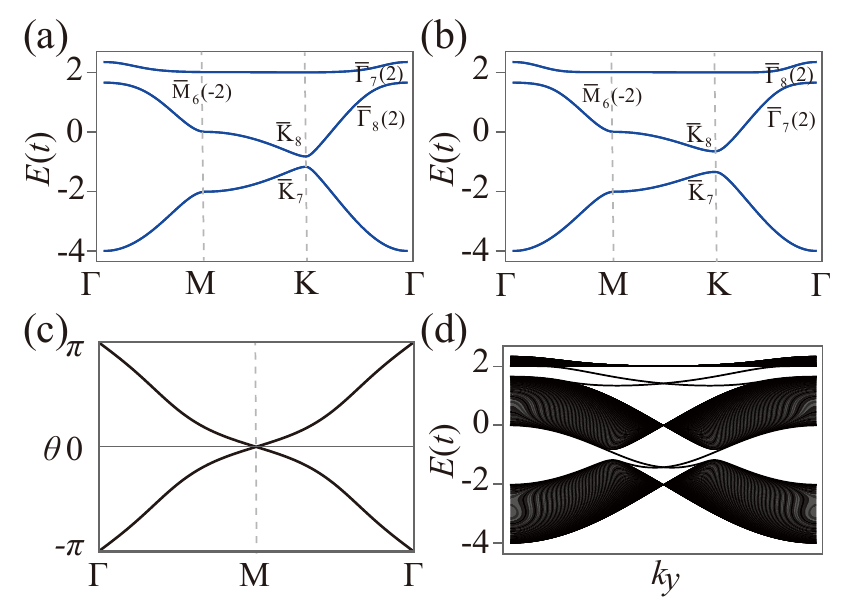}
\caption{ (a)-(b) Energy spectrum of the Kagome lattice with (a) $\lambda_{NN}=0.1~t$, $\lambda_{NNN}=0$;
(b) $\lambda_{NNN}=-0.1~t$, $\lambda_{NN}=0$. The irreps are given, with numbers in brackets indicating the character of inversion symmetry.  (c) The Wilson loop of the upper flat band with parameters $\lambda_{NN}=0.1~t$, $\lambda_{NNN}=0$. (d) The band structure of the Kagome lattice with finite-size along the $x$ direction.
\label{fig2}}
\end{figure}

In the presence of NN or NNN SOC, the band touch at the $\mathrm{\Gamma}$ point will be removed, as shown in Fig.\,\ref{fig2} (a) and (b). At the same time, Dirac cones at $K$ and $K'$ points will also become gapped. Although the upper flat band becomes weakly dispersive, we can regard it as quasi-flat when the amplitude of SOC is much smaller than $t$, as seen in Fig.\,\ref{fig2} (a) and (b), which is the case experimentally.

With SOC, the BR of the entire set of bands of Kagome lattice is  $\{\bar{\Gamma}_{7}\oplus\bar{\Gamma}_{8}\oplus\bar{\Gamma}_{9}, \bar{\mathrm{M}}_{5}\oplus2\bar{\mathrm{M}}_{6}, \bar{\mathrm{K}}_{7}\oplus\bar{\mathrm{K}}_{8}\oplus\bar{\mathrm{K}}_{9}\}$, which is an EBR $\left(\overline{E}_{g}\right)_{3f}\uparrow G$ induced from the $\overline{E}_{g}$ orbital at the Wyckoff position $3f$ of the double space group $P6/mmm$.
There are four possible ways to decompose the bands with BR $\left(\overline{E}_{g}\right)_{3f}\uparrow G$ into sets of disconnected bands, each of which satisfy the compatibility relations along all high symmetry paths, by changing the strength of SOC. These decompositions originate from the irrep pairs $\overline{\mathrm{K}}_{7}$ and $\overline{\mathrm{K}}_{8}$ ($\overline{\mathrm{\Gamma}}_{7}$ and $\overline{\mathrm{\Gamma}}_{8}$) switching partner for different amplitudes of SOC.
Indeed, when $\lambda_{NN}=0.1~t$, $\lambda_{NNN}=0$, the symmetry-data vector of the upper flat band is $\left\{\bar{\Gamma}_{7},\bar{\mathrm{M}}_{6},\bar{\mathrm{K}}_{9}\right\}$.
By contrast, the parameters $\lambda_{NN}=0$, $\lambda_{NNN}=-0.1~t$ change the symmetry-data vector to $ \left\{\bar{\Gamma}_{8},\bar{\mathrm{M}}_{6},\bar{\mathrm{K}}_{9}\right\}$.
In both of these cases, the symmetry-data vector of the flat bands is not a linear combination of EBRs where all coefficients are positive integers.
Hence, these bands are topological. In fact, the irreps $\bar{\Gamma}_{7}$ and $\bar{\Gamma}_{8}$ possess the same parity, \textit{e.g.} $+2$, and switching one of these irreps for the other does not change the index $\nu$. One can obtain $\nu=1$ from the parities of four TRIM points: $+1$ at the $\Gamma$ point and $-1$ at three $\mathrm{M}$ points.
Within the TQC theory, it is well-known that, if some sets of bands separated by a band gap, and the symmetry-data-vector of these sets of bands can be summed to a single EBR, then each set of bands possess non-trivial topology. Thus, the flat bands obtained from all of the four kinds of decompositions is topological.

Apart from the symmetry-data vector and the $\mathbb{Z}_{2}$ index $\nu$, non-trivial topology of the flat bands can also be diagnosed from the Wilson loop method and the edge-state calculation. As shown in Fig.\,\ref{fig2}(c), an odd winding number of the Wilson loop can be found, indicating $\nu=1$. By setting the strength of NN SOC as ($\lambda_{NN}=0.1~t$), we also perform the edge-state calculation of the Kagome lattice with a finite size along the $x$ direction.
As shown in Fig.\,\ref{fig2}(d), the presence of a gapless edge state between the flat band and dispersive band reflects the non-trivial topology of the flat band. In fact, for the lower dispersive band, we also have $\nu=1$. Hence, there is another gapless state between the two dispersive bands.

\section{Non-trivial Flat Bands in the Line Graph of the $S(4)$ Lattice}\label{Seciii}

\begin{figure}[t]
\includegraphics[width=8cm]{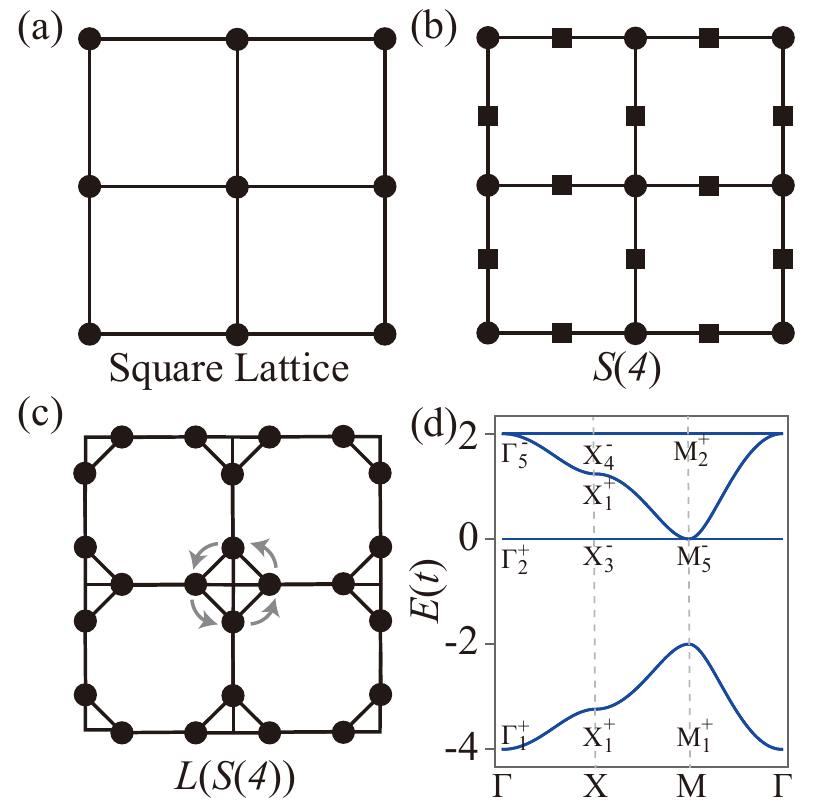}
\caption{ (a) Schematic of the square lattice. (b) The split graph of square lattice, $S(4)$. This lattice is also known as Lieb lattice.  Placing a vertex (black square) at the middle of each edge in square lattice, and considering these vertices together with the vertices and edges of square lattice, we form the split graph $S(4)$. (c) The line graph of $S(4)$.  The arrows in gray indicate that the amplitude of the considered SOC is $i\lambda$ when the spin-up electrons hop in this direction. (d) The and structure of $S(4)$ without SOC. The irreps of each band at the high-symmetry points are shown. The superscript $+/-$ denotes the parity.
\label{fig3}}
\end{figure}

In this section, we introduce the flat bands in line graphs of another kind of bipartite lattice, i.e. the split graphs $S(X)$ of bipartite lattice $X$.
As an example, the square lattice and its split graph $S(4)$ are shown in Fig.\,\ref{fig3}(a) and (b) respectively.
According to the properties given in Sec.\,\ref{Secii}, all split graphs of bipartite lattices are bipartite lattices and possess flat bands at $E=0$ and $E=2$. Moreover, the flat bands of these line graphs of $S(X)$ at $E=0$ and $E=2$ are gapless.

Following the detailed procedure in Sec.\,\ref{Secii}, in this section we identify the topologically non-trivial flat bands in SOC added line graph of the split graph of the square lattice, $L(S(4))$.

\begin{figure}[b]
\includegraphics[width=8.6cm]{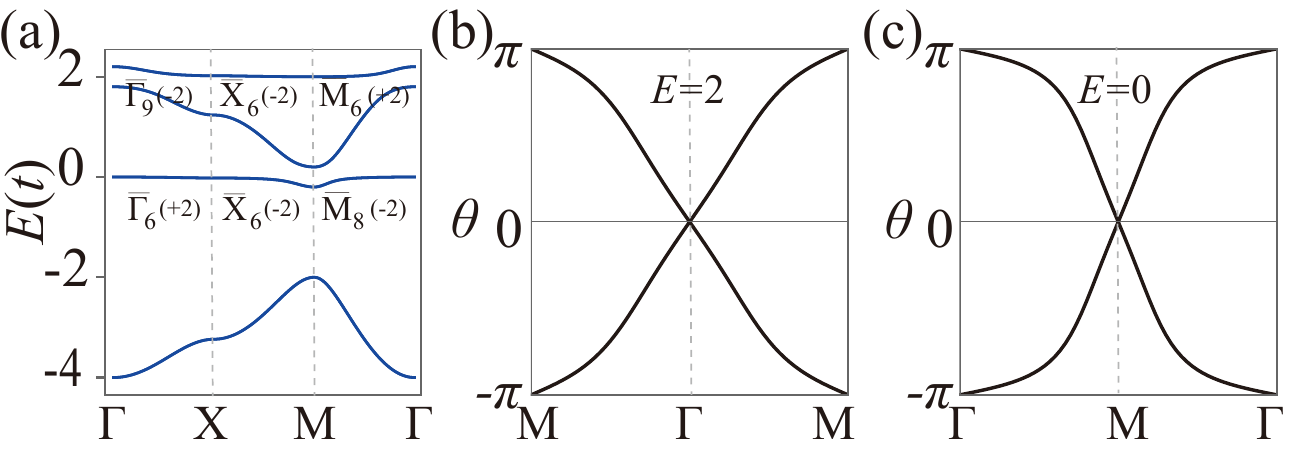}
\caption{ (a) The band structure of the line graph of $S(4)$ with nearest-neighbor SOC of amplitude $0.1~t$. There will be four separated sets of bands. The irreps are given, with numbers in brackets indicating the character of inversion symmetry. (b)-(c) The Wilson loop of the quasi-flat bands: (b) at $E=2$ and (c) at $E=0$.
\label{fig4}}
\end{figure}

\subsection{Line Graph of $S(4)$ without SOC}

The line graph of $S(4)$ is shown in Fig.\,\ref{fig3}(c), consisting of one octagon per unit cell.
As shown in Fig.\,\ref{fig3}(d), both of the flat bands located at $E=0$ and $E=2$ are ungapped due to the properties (ii) and (iv) detailed in Sec.\,\ref{Secii}.
The band touching  point is $\mathrm{\Gamma}$ ($\mathrm{M}$) for the $E=2$ ($E=0$) flat band.
Notice that the line graph of $S(4)$ preserves the symmetries of $P4/mmm$ (space group \#123) with the Wyckoff position $4l$ occupied.
The BR of the full set of bands is a linear combination of several EBRs in space group $P4/mmm$,
$\left(A_{1g}\right)_{1a}\uparrow G+\left(B_{1g}\right)_{1a}\uparrow G+\left(E_{u}\right)_{1a}\uparrow G$.
The symmetry-data vector of the upper three bands
is $\left\{\Gamma_{5}^{-}\oplus\Gamma_{2}^{+},\mathrm{M}_{5}^{-}\oplus\mathrm{M}_{2}^{+},\mathrm{X}_{1}^{+}\oplus\mathrm{X}_{3}^{-}\oplus\mathrm{X}_{4}^{-}\right\}$
which is also a linear combination of EBRs
 $\left(B_{1g}\right)_{1a}\uparrow G+\left(E_{u}\right)_{1a}\uparrow G$.
Irreps of each band at high-symmetry momenta are labeled in Fig.\,\ref{fig3} (d) and the symmetry characters of these irreps are detailed in the Table II of App.\,\ref{AppB}.
Both $\mathrm{M}_{5}^{-}$ and $\mathrm{\Gamma}_{5}^{-}$ are 2D irreps with parity $-2$.
In contrast, both $\mathrm{M}_{2}^{+}$ and $\Gamma_{2}^{+}$ are 1D irreps with parity $+1$.
As a result, as shown in Fig.\,\ref{fig3}(d), both of the flat bands at $E=0$ and $E=2$ have opposite parities at the $\mathrm{M}$ and $\Gamma$ points.
Thus, when the band touch is removed by introducing a symmetry-preserving SOC, the $\mathbb{Z}_{2}$ index becomes $\nu=1$ for each of the flat bands at $E=0$ and $E=2$.
It worth noting that, due to the $C_4$ symmetry, the $\mathbb{Z}_{2}$ index is independent of the parity at the $\mathrm{X}$ point.

\subsection{Line Graph of $S(4)$ with SOC}
Upon adding NN SOC with strength $0.1t$, we calculate the band structures of the line graph of $S(4)$.
As shown in Fig.\,\ref{fig4} (a), both of the band touches at the $\Gamma$ and $\mathrm{M}$ points are simultaneously gapped by the SOC.
The  symmetry-data vector of the quasi-flat band near $E=2$ ($E=0$) is $ \left\{\overline{\Gamma}_{9}, \overline{\mathrm{X}}_{6}, \overline{\mathrm{M}}_{6} \right\}$ ($ \left\{\overline{\Gamma}_{6},\overline{\mathrm{X}}_{6},\overline{\mathrm{M}}_{8} \right\}$).
The symmetry-data vector of the dispersive band between the two quasi-flat bands is $ \left\{ \overline{\Gamma}_{8},\overline{\mathrm{X}}_{5},\overline{\mathrm{M}}_{9}\right\}$.
As the parity of the irreps $\overline{\Gamma}_{8}$, $\overline{\Gamma}_{9}$, $\overline{\mathrm{M}}_{8}$ and $\overline{\mathrm{M}}_{9}$ are the same and equal to $-2$, the topological index $\nu=1$ does not change by swapping the order of bands with irreps $\overline{\Gamma}_{8}$ and $\overline{\Gamma}_{9}$ or $\overline{\mathrm{M}}_{8}$ and $\overline{\mathrm{M}}_{9}$.

The Wilson loop calculations of the quasi-flat bands at $E=2$ and $E=0$ are shown in Fig.\,\ref{fig4} (b) and (c), respectively, where the topologically non-trivial phase is indicated by an odd winding number of the Wilson loop.

\section{Discussion and Conclusion}\label{Secv}

In Sec.\,\ref{Secii} and Sec.\,\ref{Seciii}, we find that with SOC, the quasi-flat bands in the Kagome lattice and $L(S(4))$ lattices are topologically nontrivial.
In both cases, the degenerate point forms a 2D band representation with character of the inversion symmetry equal to $2$ or $-2$.
Thus, any kind of symmetry-allowed SOC cannot change the $\mathbb{Z}_{2}$ index, and induce a set of flat bands to be topological when the band touch is removed.
We also explore the line graph of square lattice $L(4)$ and the line graph of the split lattice of the honeycomb lattice $L(S(6))$, see App.\,\ref{AppC} for details.
Similar to that of the Kagome and $L(S(4))$ lattices, the band touches in the $L(4)$ and $L(S(6))$ lattices form 2D irreps.
Quasi-flat bands result from adding SOC, and they are also topologically nontrivial with $\mathbb{Z}_{2}$ index $\nu=1$.
Apart from line graphs which contain only one flat band (without SOC), one can also find bipartite lattices whose line graphs have flat-band degeneracies $D>1$.
Such root graph lattices include the octagon-square lattice and the hexagon-square lattice, which are studied in App.\,\ref{AppD}.
With the nearest-neighbor SOC taken into consideration, we find the resulting set of quasi-flat bands are also topologically nontrivial.

Split graphs of bipartite lattices comprise an entire class of lattices with flat bands.
In contrast to the flat bands of the line graph lattices considered, the flat bands in these split graphs are at $E=0$.
We discuss the topological properties of these flat bands Apps.\,\ref{AppE} and \,\ref{AppF}, where we find both strong topological and fragile topological states.

In summary, following the TQC theory, we investigate the topological properties of flat bands in line graphs and split graphs of bipartite lattices with symmetry-allowed SOC.
For the line graph of $X$, there is a set of flat bands at $E=2$, which becomes topologically nontrivial once the band touch is removed.
For the line graph of $S(X)$, there are two sets of flat bands, one each at $E=0$ and $E=2$.
Both of these sets of flat bands become topologically nontrivial when SOC is added.
Finally, for split graphs of bipartite lattices, we find that adding Rashba SOC results in quasi-flat bands which are strong topological.
Our results provide a generic way to obtain flat bands with non-trivial topology, as a path to explore strongly interacting systems.

\begin{acknowledgements}
B.A.B thanks  R.J. Cava, N.P. Ong and A. Yazdani for discussions. This work was supported by the DOE Grant No. DE-SC0016239, the Schmidt Fund for Innovative Research, Simons Investigator Grant No. 404513, and the Packard Foundation. Further support was provided by the NSF-EAGER No. DMR 1643312, NSF-MRSEC No. DMR-1420541, ONR No. N00014-20-1-2303, MURI W911NF-15-1-0397 , Gordon and Betty Moore Foundation through Grant GBMF8685 towards the Princeton theory program, BSF Israel US foundation No. 2018226, and the Princeton Global Network Funds. Y.X. and B.A.B. were also supported by the Max Planck society.
\end{acknowledgements}

\begin{appendix}

\section{Properties of bipartite lattices, line graphs, and split graphs}\label{AppA}
Here, we prove the properties that are relevant to our work. Let $X$ be a bipartite root graph with $m$ polygons per unit cell.

\noindent \textbf{(i). A graph $X$ is a bipartite lattice if and only if all the polygons from this graph are even-sided.}

Following the definition of a bipartite lattice, vertices in any closed loop of the lattice can be divided into two sets of the same size. Thus, the total number of vertices in the closed loop must be even, indicating that the closed loop is an even-sided polygon. In addition, if the root graph $X$ consists entirely of even-sided polygons, then vertices along each polygon can be divided into the subsets $U$ and $V$ without any contradiction, establishing the graph to be bipartite.

\noindent \textbf{(ii). There is always at least one flat band at $E=-2t$ touching the dispersive bands in the energy spectrum of the line graph $L(X)$ of $X$.}

We will show that this is the case through constructing the states which span the flat band.
The bipartite lattice $X$ is formed by even-sided polygons.
Notice that the line graph of $X$ always possesses the same number of such even-sided polygons.
Each even-sided polygon supports a cycle-like compact localized state (CLS) \cite{mielke1991ferromagnetic,PhysRevB.78.125104,kollar2019line}.
Define a wavefunction whose amplitude at any vertex in the closed loop is real-valued and equal to $1$ or $-1$ (unnormalized), and $0$ for any vertex not in the loop.
Then every site of amplitude $+1$ $(-1)$ has two neighbors of amplitude $-1$ $(+1)$.
The resulting CLS is thus an eigenfunction of eigenvalue exactly $-2t$.
As one of these CLSes can be defined for every polygon, these states span the flat band at $E=-2t$.
When $L(X)$ is embedded on a torus with $N$ unit cells (\textit{i.e.} the periodic boundary condition is applied), one finds that all of the $mN$ CLSes will sum to zero (given a certain phase per CLS). One can additionally find two linearly independent extended real-space eigenstates. Then, one has $mN+1$ real-space eigenstates in total, $mN$ of which fill the band, necessarily resulting in a band touch \cite{PhysRevB.78.125104}.

\noindent \textbf{(iii). The degeneracy $D$ of the flat bands in $L(X)$ is given by $m$.}

Based on property \textbf{(ii)}, each even-sided polygon in $L(X)$ can support a CLS with $E=-2t$. Thus, the degeneracy of the flat bands is $D=m$.

\noindent \textbf{(iv). The split graph $S(X)$ is always bipartite.}

If a polygon in the root graph has $n$ vertices, the corresponding loop in $S(X)$ will have $2n$ vertices.

\noindent \textbf{(v). There are a set of flat bands with degeneracy $D=m$ located at $E=0$ in the spectra of $S(X)$ and $L(S(X))$.}

Similarly to the CLS construction for $L(X)$, if a loop has $4k$ vertices, where $k$ is a integer, one can construct a CLS with eigenvalue $0$ along this loop in $S(X)$ and $L(S(X))$ \cite{kollar2019line}. As each polygon in $S(X)$ and $L(S(X))$ has $4k$ vertices and thus supports such a CLS, the degeneracy is $D=m$.

This construction holds as long as the bipartite vertex subsets $U$ and $V$ are not of the same size \cite{PhysRevLett.62.1201}.
For split graphs of bipartite lattices, the subset $U$ can be defined to be the set of all vertices in the bipartite lattice, while the subset $V$ can be defined to be the set of all newly introduced vertices, necessarily of size equal to the number of edges in the bipartite lattice.
Therefore, as long as the number of vertices and edges in the original bipartite lattice are unequal, which is the case for any non-cycle graph, the split graph of the bipartite lattice will have at least one flat band at energy $E=0$.

\noindent \textbf{(vi). The flat bands in the energy spectra of $L(X)$, $S(X)$, and $L(S(X))$ are always ungapped with a set of dispersive bands.}

This proof closely follows that of property (ii).

\section{Character Tables of Irreps Mentioned}\label{AppB}

Within the TQC framework, one can diagnose the band topology of a set of isolated bands by the symmetry properties of these bands. The symmetry property is described by its decompositions into irreps of little groups at the maximal momenta in the first  Brillouin Zone. The concept of symmetry-data vector $B$ is defined in TQC to characterize the symmetry properties, and its explicit form is
\begin{eqnarray}
B&=&\left\{ n_{1}^1G_{K_{1}}^{1}\oplus n_1^2G_{K_{1}}^{2}\oplus \cdots,n_2^1G_{K_{2}}^{1}\oplus n_2^2G_{K_{2}}^{2}\oplus \cdots\right\},
\end{eqnarray}
where $G_{K_{i}}^{j}$ is the $j$-th irrep of the little group at the  maximal momenta $K_{i}$, and the $n_i^j$ is the multiplicity of irrep $G_{K_{i}}^{j}$.
If the symmetry-data vector of any set of gapped bands cannot be written as a sum of EBRs, it must be topologically non-trivial.

Here, we tabulate the character tables of irreps that form the symmetry-data vectors of flat bands in the lattices studied in this paper. The irreps of little groups of space group $P6/mmm$, $P4/mmm$ and $P2/m$ are tabulated in Tables \ref{tb1}, \ref{tb2}, and \ref{tb3}, respectively.

\begin{table}[h]
\caption{Character table of irreps of little groups at the high symmetry momenta $\Gamma: (0,0,0)$, $\mathrm{K}: (1/3,1/3,0)$, and $\mathrm{M}: (1/2,0,0)$, of the space group $P6/mmm$ with time-reversal symmetry. For the little group at the $\Gamma$ point, $E$, $C_3$, $C_2$, $C_6$, $C_{2}^{'}$ and $C_{2}^{''}$ represent the conjugation classes generated from the Identity, $C_{3z}$, $C_{2z}$, $C_{6z}$, $C_{2y}$ and $C_{2x}$. $ \overline{E}$, $\overline{C_{3}}$, $\overline{C_{2}}$, $\overline{C_{6}}$, $\overline{C_{2}^{'}}$, $\overline{C_{2}^{''}}$ represent the conjugation classes generated from the symmetry combining inversion symmetry with the Identity, $C_{3z}$, $C_{2z}$, $C_{6z}$, $C_{2y}$, and $C_{2x}$ respectively. Conjugate class symbols at the $\mathrm{K}$ and $\mathrm{M}$ points are defined in a similar manner.
}
\vspace{0.2cm}
\renewcommand\arraystretch{1.5}
\begin{tabular}{p{0.6cm}<{ \centering}p{0.5cm}<{\centering}p{0.5cm}<{\centering}p{0.5cm}<{\centering}p{0.5cm}<{\centering}p{0.5cm}<{\centering}p{0.5cm}<{\centering}p{0.5cm}<{\centering}p{0.5cm
}<{\centering}p{0.5cm}<{\centering}p{0.5cm}<{\centering}p{0.5cm}<{\centering}p{0.5cm}<{\centering}p{0.5cm}}
\hline
\hline
   & $ E $  & $ 2C_{3}$ &$ C_{2}$ &$2C_{6}$ &$3C_{2}^{'}$ &$3C_{2}^{''}	$ &$ \overline{E}$ &$2\overline{C_{3}}$ &$\overline{C_{2}}$ &$2\overline{C_{6}}$ &$3\overline{C_{2}^{'}}$ &$3\overline{C_{2}^{''}}$ \\
\hline
 $\Gamma_{1}^{+}$  & 1 & 1  & 1   & 1 & 1  & 1  & 1   & 1 & 1  & 1 & 1 & 1      \\
 $\Gamma_{5}^{+}$  & 2 & -1  & 2   & -1 & 0  & 0  & 2   & -1 & 2  & -1 & 0 & 0      \\
 $\Gamma_{6}^{-}$  & 2 & -1  & -2   & 1 & 0  & 0  & -2   & 1 & 2  & -1 & 0 & 0      \\
\hline
 &$E$&$2C_{3}$&$3C_{2}^{'}$&$\overline{C_{2}}$&$2\overline{C_{6}}$&$3\overline{C_{2}^{''}}$\\
 \hline
$\mathrm{K}_{1}$  & 1 & 1  & 1   & 1 & 1  & 1      \\
$\mathrm{K}_{4}$  & 1 & 1  & -1   & 1 & 1  & -1      \\
$\mathrm{K}_{5}$  & 2 & -1  & 0   & 2 & -1  & 0      \\
\hline
&$E$&$C_{2}$&$C_{2}^{'}$&$C_{2}^{''}$&$\overline{E}$&$\overline{C_{2}}$&$3\overline{C_{2}^{'}}$&$3\overline{C_{2}^{''}}$ \\
\hline
$\mathrm{M}_{1}^{+}$&1	&1	&1	&1	&1	&1	&1	&1\\
$\mathrm{M}_{2}^{+}$&1	&1	&-1	&-1	&1	&1	&-1	&-1\\
$\mathrm{M}_{3}^{-}$&1	&-1	&1	&-1	&-1	&1	&-1	&1\\
$\mathrm{M}_{4}^{-}$&1	&-1	&-1	&1	&-1	&1	&1	&-1\\
\hline
\hline
\end{tabular}\label{tb1}
\end{table}

\begin{table}[h]
\caption{Character table of irreps of little groups at the high symmetry momenta $\Gamma: (0,0,0)$, $\mathrm{X}: (0,1/2,0)$, and $\mathrm{M}: (1/2,1/2,0)$, of the space group $P4/mmm$ with time-reversal symmetry. For the little group at the $\Gamma$ point, $E$, $C_2$, $C_4$, $C_{2}^{'}$ and $C_{2}^{''}$ represent the conjugation classes generated from the Identity, $C_{2z}$, $C_{4z}$,$C_{2y}$ and $C_{2x}$. $ \overline{E}$, $\overline{C_{2}}$, $\overline{C_{4}}$,  $\overline{C_{2}^{'}}$, $\overline{C_{2}^{''}}$ represent the conjugation classes generated from the symmetry combining inversion symmetry with the Identity, $C_{3z}$, $C_{2z}$, $C_{4z}$, $C_{2y}$, and $C_{2x}$ respectively. Conjugate class symbols at the $\mathrm{X}$ and $\mathrm{M}$ points are defined in a similar manner.
}
\vspace{0.2cm}
\renewcommand\arraystretch{1.5}
\begin{tabular}{p{0.6cm}<{ \centering}p{0.6cm}<{\centering}p{0.6cm}<{\centering}p{0.6cm}<{\centering}p{0.6cm}<{\centering}p{0.6cm}<{\centering}p{0.6cm
}<{\centering}p{0.6cm}<{\centering}p{0.6cm}<{\centering}p{0.6cm}<{\centering}p{0.6cm}<{\centering}p{0.6cm}}
\hline
\hline
   & $ E $  & $ C_{2}$ &$ 2C_{4}$ &$2C_{2}^{'}$ &$2C_{2}^{''}	$ &$ \overline{E}$ &$\overline{C_{2}}$ &$2\overline{C_{4}}$ &$2\overline{C_{2}^{'}}$&$2\overline{C_{2}^{''}}$ \\
\hline
 $\Gamma_{1}^{+}$  & 1 & 1  & 1   & 1 &  1  & 1 & 1  & 1   & 1   & 1        \\
 $\Gamma_{2}^{+}$  & 1 & 1  & -1   & 1 &  -1  & 1 & 1  & -1   & 1   & -1        \\
 $\Gamma_{4}^{+}$  & 1 & 1  & -1   & -1 &  1  & 1 & 1  & -1   & -1 &  1         \\
 $\Gamma_{5}^{-}$  & 2 & -2  & 0   & 0 & 0  & -2  & 2   & 0 &0  & 0      \\
\hline
   & $ E $  & $ C_{2}$ &$C_{2}^{'}$&$C_{2}^{''}$  &$ \overline{E}$ &$\overline{C_{2}}$ &$\overline{C_{2}^{'}}$&$\overline{C_{2}^{''}}$ \\
 \hline
$\mathrm{X}_{1}^{+}$  & 1 & 1  & 1   & 1 & 1  & 1 & 1 & 1      \\
$\mathrm{X}_{3}^{-}$  & 1 & -1  & -1   & 1 & -1  & 1 & 1 & -1     \\
$\mathrm{X}_{4}^{-}$  & 1 & -1  & 1   & -1 & -1  & 1 & -1 & 1     \\
\hline
& $ E $  & $ C_{2}$ &$ 2C_{4}$ &$2C_{2}^{'}$ &$2C_{2}^{''}	$ &$ \overline{E}$ &$\overline{C_{2}}$ &$2\overline{C_{4}}$ &$2\overline{C_{2}^{'}}$&$2\overline{C_{2}^{''}}$ \\
\hline
 $\mathrm{M}_{1}^{+}$  & 1 & 1  & 1   & 1 &  1  & 1 & 1  & 1   & 1   & 1        \\
 $\mathrm{M}_{2}^{+}$  & 1 & 1  & -1   & 1 &  -1  & 1 & 1  & -1   & 1   & -1        \\
 $\mathrm{M}_{4}^{+}$  & 1 & 1  & -1   & -1 &  1  & 1 & 1  & -1   & -1 &  1         \\
 $\mathrm{M}_{5}^{-}$  & 2 & -2  & 0   & 0 & 0  & -2  & 2   & 0 &0  & 0      \\
\hline
\hline
\end{tabular}\label{tb2}
\end{table}

\begin{table}[h]
\caption{Character table of irreps of little groups at the high symmetry momentum $\Gamma: (0,0,0)$, of the space group $P2/m$ with time-reversal symmetry. The character of irrep $O_{i}^{+/-}$ is equal to that of the $\Gamma_{i}^{+/-}$ (with $i=1,2$ and $O=\mathrm{X}$, $\mathrm{Y}$ or $\mathrm{M}$) and represents the irerp of the little group at the momentum $\mathrm{X}: (1/2,0,0)$, $\mathrm{Y}: (0,1/2,0)$, and $\mathrm{M}: (1/2,1/2,0)$
}
\vspace{0.2cm}
\renewcommand\arraystretch{1.5}
\begin{tabular}{p{0.6cm}<{ \centering}p{0.6cm}<{\centering}p{0.6cm}<{\centering}p{0.6cm}<{\centering}p{0.6cm}<{\centering}p{0.6cm}<{\centering}p{0.6cm
}<{\centering}p{0.6cm}<{\centering}p{0.6cm}<{\centering}p{0.6cm}<{\centering}p{0.6cm}<{\centering}p{0.6cm}}
\hline
\hline
   & $ E $  & $ C_{2}$ &$ \overline{E}$ &$\overline{C_{2}}$ \\
\hline
 $\Gamma_{1}^{+}$  & 1 & 1  & 1   & 1      \\
 $\Gamma_{1}^{-}$  & 1 & 1  & -1   & -1       \\
 $\Gamma_{2}^{+}$  & 1 & -1  & 1   & -1      \\
 $\Gamma_{2}^{-}$  & 1 & -1  & -1   & 1    \\
\hline
\hline
\end{tabular}\label{tb3}
\end{table}

\section{Additional Examples}\label{AppC}

\begin{figure}[t]
\includegraphics[width=8.6cm]{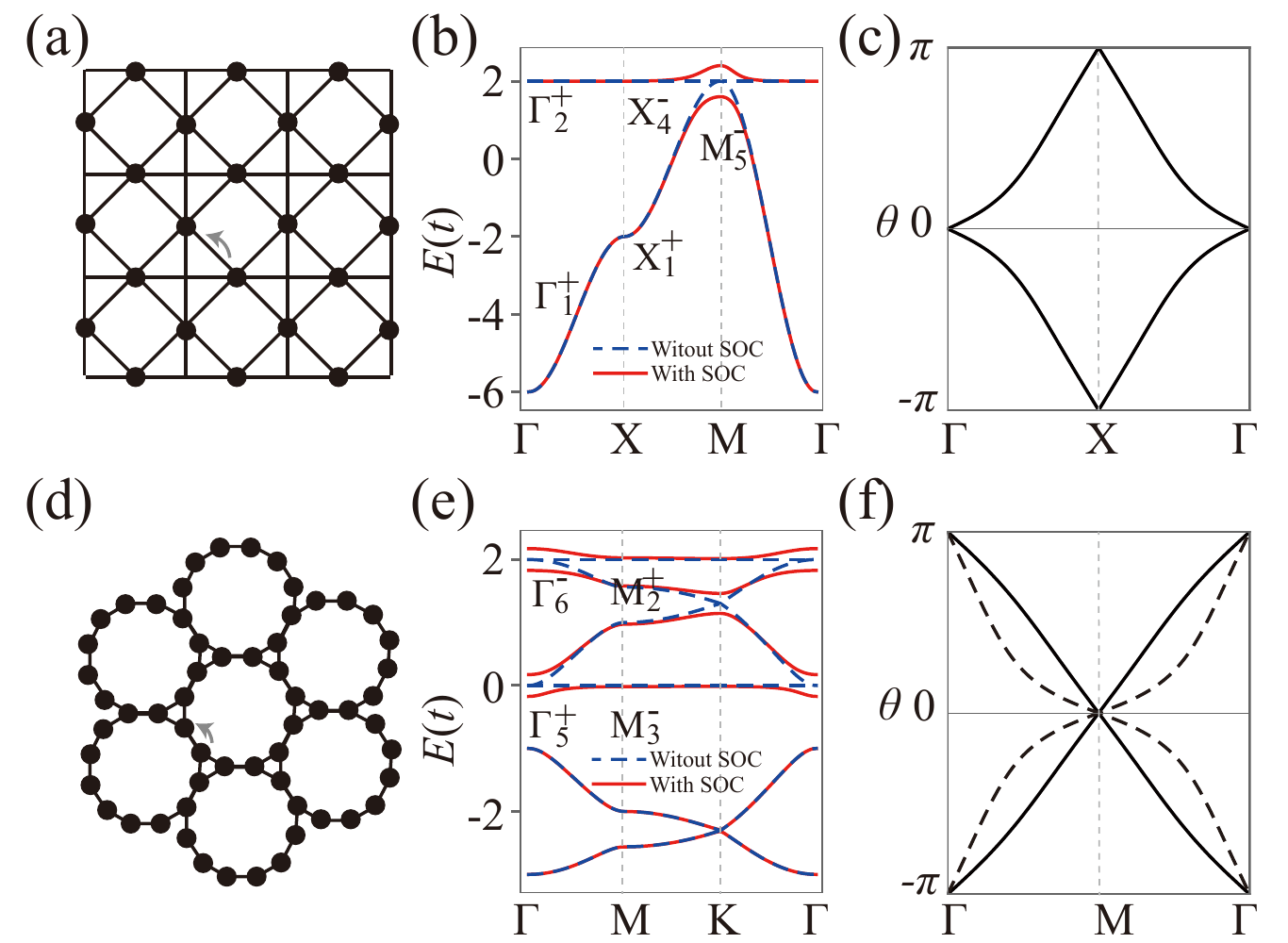}
\caption{ (Color  online)  (a)  Schematic of the line graph of the square lattice. The arrow in gray means the amplitude of the considered SOC is $i\lambda$ when the up-spin atoms hop along the arrowed direction. (b) The band structure of line graph of square lattice without SOC (dashed blue line) and with SOC (red line). The irreps of the flat band without SOC at the high symmetry points are shown. The superscript $+/-$ means the parity. The amplitude of SOC set as $0.1~t$. (c) The Wilson loop of the quasi-flat bands of line graph of square lattice with SOC. (d)-(f) Corresponding plots starting from the line graph of the split graph of the honeycomb lattice. There are two sets of flat bands. The Wilson loop of the upper (middle) flat band is shown in (f) by black (dashed-black) line.
\label{fig5}}
\end{figure}

Line graphs and split graphs can be used to construct an entire class of lattices with a single flat band in the absence of SOC. Here, we study the topologically non-trivial quasi-flat bands in the SOC-added line graph of the square lattice and the SOC-added line graph of $S(6)$ (the split graph of the honeycomb lattice). Both of these lattices are shown in Fig.\,\ref{fig5}.

\noindent\textbf{Line Graph of the Square Lattice:}

The line graph of the square lattice is shown in Fig.\,\ref{fig5} (a).
This lattice possesses one flat band and one dispersive band, and these two bands have a band touch at the $\mathrm{M}$ point as shown in Fig.\,\ref{fig5}(b).
This lattice belongs to the $P4/mmm$ space group.
The symmetry-data vector of the two bands in the spinless case is  $\left\{\Gamma_{1}^{+}\oplus\Gamma_{2}^{+},\mathrm{M}_{5}^{-},\mathrm{X}_{1}^{+}\oplus\mathrm{X}_{4}^{-}\right\}$, which is an EBR $\left(A_{g}\right)_{2f}\uparrow G$ induced by an $s$ orbital centered on $2f$.
The parity of the flat band at the $\Gamma$ point is $1$, while at the $\mathrm{M}$ point it is $-1$ .
Thus, once a gap is introduced by SOC, the quasi-flat band will be topologically non-trivial with $\nu =1$.

The energy spectrum of the line graph of the square lattice with NN SOC is shown in Fig.\,\ref{fig5}(b).
The bands become gapped at the $\mathrm{M}$ point.
The symmetry-data vector of the full set of four bands is a decomposable single EBR: $\left(\bar{E}_{g}\right)_{2f}\uparrow G$.
Thus, both sets of bands in this lattice are topologically non-trivial.
The band topology of the quasi-flat bands is further diagnosed by the Wilson loop as shown in Fig.\,\ref{fig5}(c), where one can find an odd Wilson loop winding.

\noindent\textbf{Line Graph of the $S(6)$ Lattice:}

The $S(6)$ lattice is the split graph of the honeycomb lattice.
The line graph of $S(6)$ is shown in Fig.\,\ref{fig5}(d).
In this lattice, the (non-maximal) Wyckoff position $6l$ of the space group $P6/mmm$ is occupied.
As mentioned in Section \ref{Seciii}, the spectrum of all line graphs of split graphs of bipartite lattices consist of two sets of flat bands and some dispersive bands.
As shown in Fig.\,\ref{fig5}(e), both of the $E=2$ and $E=0$ flat bands touch dispersive bands at the $\Gamma$ point.
In the spinless case, the energy spectrum of this lattice can be divided into two branches: the upper four bands and the lower two bands.
The BR of the full set of bands is a sum of EBRs, given by $\left(A_1'\right)_{2c}\uparrow G \oplus \left(E'\right)_{2c}\uparrow G$.
The symmetry-data vector of the upper four bands is a single EBR induced by an $E'$ orbital centered on $2c$ and reads as
$\left\{\Gamma_{5}^{+}\oplus\Gamma_{6}^{-},\mathrm{K}_{1}\oplus\mathrm{K}_{4}\oplus\mathrm{K}_{5},\mathrm{M}_{1}^{+}\oplus\mathrm{M}_{2}^{+}\oplus\mathrm{M}_{3}^{-}\oplus\mathrm{M}_{4}^{-}\right\}$. The irreps of the $E=2$ ($E=0$) flat band at the $\mathrm{M}$ point is $\mathrm{M}_{2}^{+}$ ($\mathrm{M}_{3}^{-}$), and the irrep of the $E=2$ ($E=0$) band touch is $\Gamma_{6}^{-}$ ($\Gamma_{5}^{+}$).
Thus, both of the flat bands will be topologically non-trivial with $\nu=1$ when the band touch is removed by SOC.

The band structure of the line graph of $S(6)$ with NN SOC is shown in Fig.\,\ref{fig5}(e).
Both band touches will be removed by SOC. The Wilson loops of the gapped quasi-flat bands are shown in Fig.\,\ref{fig5}(f). The odd Wilson loop winding indicates non-trivial band topology.

\begin{figure}[t]
\includegraphics[width=8.6cm]{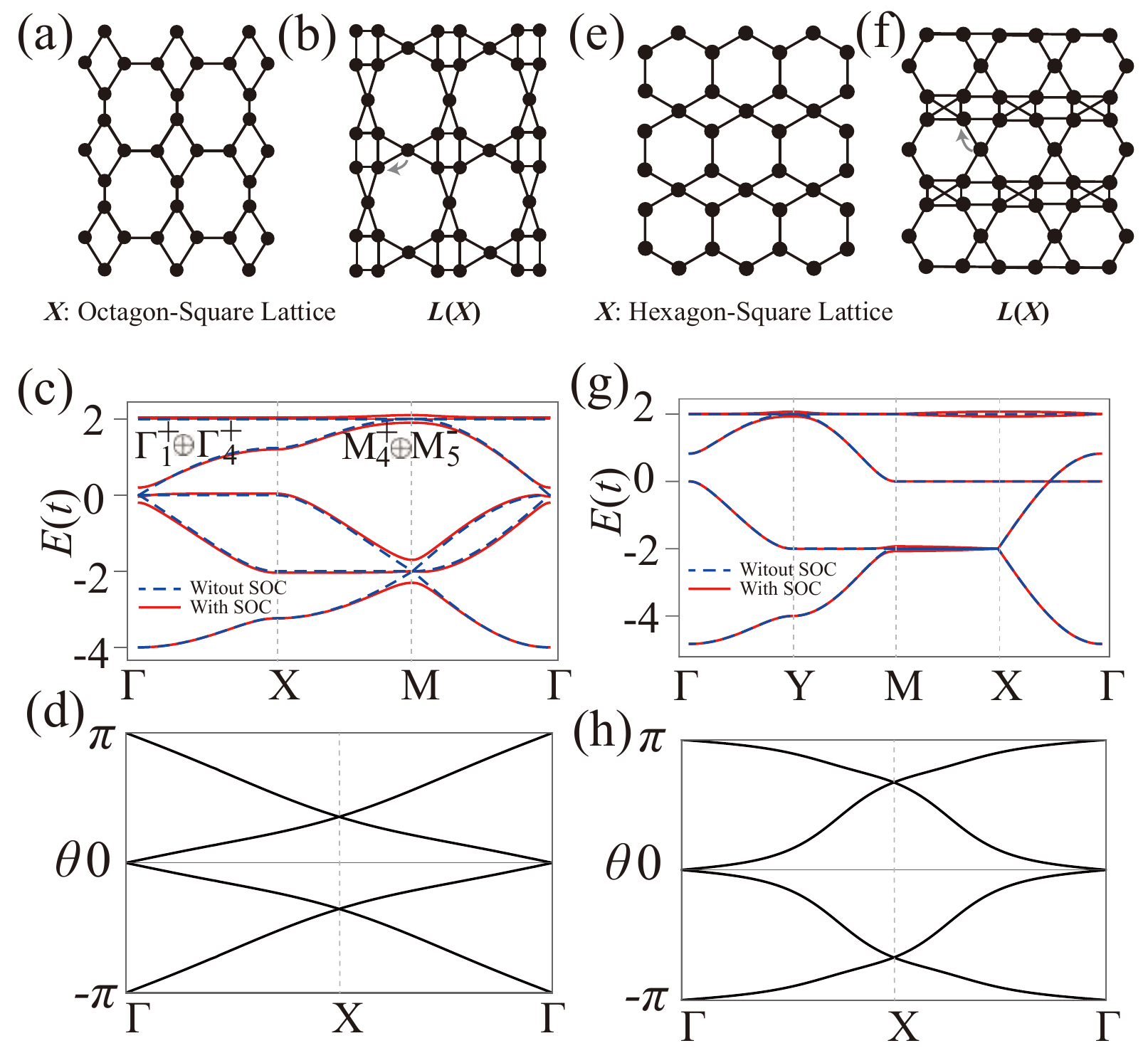}
\caption{(Color  online)  (a)  Schematic picture of the octagon-square lattice. (b) The line graph of octagon-square lattice. The arrow in gray means the amplitude of the considered SOC is $i\lambda$ when the up-spin atoms hop along the arrowed direction.  (c) The band structure of line graph of octagon-square lattice without SOC (dashed blue line) and with SOC (red line). Here, the amplitude of SOC set as $0.1~t$. (d) The Wilson loop of the two upper sets of quasi-flat bands of line graph of octagon-square lattice with SOC. (e)-(h) Corresponding plots starting from the hexagon-square lattice.
\label{fig6}}
\end{figure}

 \begin{table*}[t]
\caption{parity of the top three bands (six bands in spinful case) of the line graph of hexagon-square lattice at the high symmetry points.
}
\vspace{0.2cm}
\renewcommand\arraystretch{1.5}
\begin{tabular}{p{2.6cm}<{ \centering}p{2.6cm}<{\centering}p{2.6cm}<{\centering}p{2.6cm}<{\centering}p{2.6cm}<{\centering}p{2.6cm}<{ \centering}}
\hline
\hline
 &  & $\Gamma$ & $\mathrm{X}$ & $\mathrm{Y}$ & $\mathrm{M}$\tabularnewline
\hline
spinless & flat+dispersive bands & $\left(-1,1,1\right)$ & $\left(-1,-1,1\right)$ & $\left(-1,1,1\right)$ & $\left(-1,1,1\right)$\tabularnewline
\hline
\multirow{2}{*}{spinful} & flat bands & $\left(-1,-1,1,1\right)$ & $\left(-1,-1,-1,-1\right)$ & $\left(1,1,-1,-1\right)$ & $\left(-1,-1,1,1\right)$\tabularnewline
 & dispersive bands & $\left(1,1\right)$ & $\left(1,1\right)$ & $\left(1,1\right)$ & $\left(1,1\right)$\tabularnewline
\hline
\hline
\end{tabular}\label{tb4}
\end{table*}

\section{Examples with Higher Degeneracy}\label{AppD}

We have identified that the flat band in the line graph of bipartite lattices with $D=1$ will be topologically non-trivial once the band touch is removed by adding SOC.
Here, we turn to line graphs of bipartite lattices with a greater number of polygons per unit cell, where the degeneracy of flat bands is higher.

\noindent\textbf{Line Graph of Octagon-Square Lattice:}

As schematically shown in Fig.\,\ref{fig6}(a), the octagon-square lattice belongs to the space group $P4/mmm$ and possesses one octagon and one square per unit cell.
The line graph of this lattice is shown in Fig.\,\ref{fig6}(b).
Then, because there are two polygons per unit cell, the flat band is doubly degenerate in the absence of SOC.
As shown in Fig.\,\ref{fig6}(c), the double-degenerate flat bands touch the dispersive band at the $\mathrm{M}$ point, forming a triple-degenerate point.
Due to the $C_4$ symmetry, the $\mathbb{Z}_{2}$ index is independent of the parity at the $\mathrm{X}$ point. As such, we focus on the parity at the $\Gamma$ and $\mathrm{M}$ points.
The irrep of the double-degenerate flat bands at the $\Gamma$ point is $\Gamma_{1}^{+}\oplus\Gamma_{4}^{+}$ with parity equal to $+1$.
The irrep at the $\mathrm{M}$ point of the triple-degeneracy point is $\mathrm{M}_{4}^{+}\oplus\mathrm{M}_{5}^{-}$.
Here, $\mathrm{M}_{4}^{+}$ is 1D irrep with parity $1$, while $\mathrm{M}_{5}^{-}$ is a 2D irrep with parity $-2$.

We then introduce nearest-neighbor SOC as schematically shown in Fig.\,\ref{fig6}(b).
The degeneracy of the flat bands will be broken, leaving a set of two quasi-flat bands, see in Fig.\,\ref{fig6}(c).
Moreover, the triple-degenerate point is split into three sets of isolated bands (six bands in total) at the $\mathrm{M}$ point.
From the irreps obtained in the spinless case, we know that the parity of one of these three sets of bands is $1$, leaving parity $-1$ for the other two sets of bands.
With the addition of NN SOC, we find that the parities of the two sets of quasi-flat bands at the $\mathrm{M}$ point are $\left(-1,1\right)$, while that of the set of dispersive bands is $-1$.
The parities of both quasi-flat bands at the $\Gamma$ point are $1$.
Then, one finds that the $\mathbb{Z}_{2}$ index of the quasi-flat bands is $1$. The Wilson loop of the upper two sets of quasi-flat bands is shown in Fig.\,\ref{fig6}(d), indicating odd Wilson loop winding.

\noindent\textbf{Line Graph of Hexagon-Square Lattice:}

Similar to the octagon-square lattice, the hexagon-square lattice is formed by one hexagon and one square per unit cell, shown in Fig.\,\ref{fig6}(e).
The line graph of the hexagon-square lattice belongs to the space group $P2/m$ and is shown in Fig.\,\ref{fig6}(f).
In the spinless case, the flat bands are double-degenerate and touch one dispersive band at the $\mathrm{Y}$ point, see Fig.\,\ref{fig6}(g).
The irreps of the double-degenerate flat bands are $\Gamma_{1}^{+}\oplus\Gamma_{2}^{-}$ at the $\Gamma$ point, $2\mathrm{X}_{2}^{-}$ at the $\mathrm{X}$ point, and $\mathrm{M}_{1}^{+}\oplus\mathrm{M}_{2}^{-}$ at the $\mathrm{M}$ point.
The character tables of these irreps can be found in App.\,\ref{AppB}, and the parity is given by the superscript of the symbol of each irrep.
At the $\mathrm{Y}$ point, the little group irrep formed by the triple-degenerate cone is $2\mathrm{Y}_{1}^{+}\oplus\mathrm{Y}_{2}^{-}$.
The parities at the high-symmetry points are given in Table \ref{tb4}.

With NN SOC added, the bands will become gapped at the $\mathrm{Y}$ point as shown in  Fig.\,\ref{fig6}(g). The parity of the set of dispersive bands at $\mathrm{Y}$ is $1$, and that of the two sets of quasi-flat bands are $\left(1,-1\right)$, see Table \ref{tb4}. With the parities of the quasi-flat bands at the other TRIMs, tabulated in Table. \ref{tb4}, we have that $\nu=1$. The Wilson loop exhibits an odd winding number of the upper two sets of quasi-flat bands and is shown in Fig.\,\ref{fig6}(h).

\section{Nontrivial Flat Bands in Split Graphs of Bipartite Lattices}\label{AppE}
\begin{figure*}[t]
\includegraphics[width=17cm]{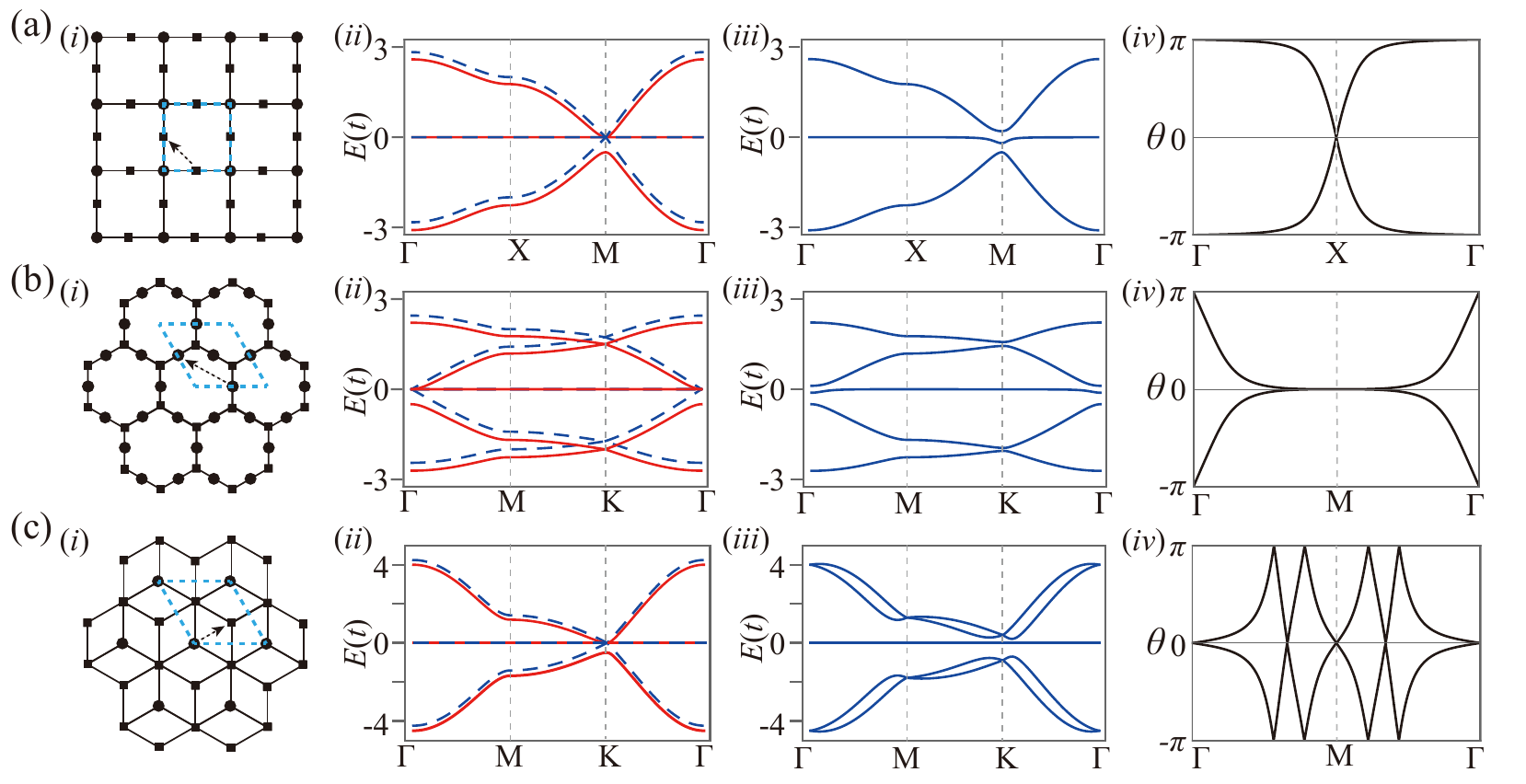}
\caption{(a) (\textit{i}) Schematic picture of the split graph of the square lattice, $S(4)$. The unit cell is highlighted. Atoms can be divided into two sets $U$ and $V$ as shown by the black dots and squares. The degeneracy of the flat band is given by the difference between the numbers of atoms belonging to the two sets. The SOC amplitude is $i\lambda$ for spin-up electrons hopping along the direction of the black dotted arrow, with $\lambda=0.2~t$. (\textit{ii}) The band structure of $S(4)$. The band structure without (with) on-site energy is highlighted in dashed blue (solid red). The on-site energy is set to $0.5~t$. (\textit{iii}) Band structure of the $S(4)$ lattice with SOC. (\textit{iv}) The Wilson loop of the quasi-flat band.
(b)-(c) Corresponding plots for the split graph of the honeycomb lattice and the dice lattice. Rashba SOC in $S(4)$ and $S(6)$ preserves inversion symmetry, while in the dice lattice preserves $C_6$ symmetry.
\label{fig7}}
\end{figure*}

The split lattice $S(X)$ is well-known to have flat bands at $E=0$.
These flat bands arise because split graphs are bipartite graphs where the number of vertices in subsets $U$ and $V$ (as defined in Sec.\,\ref{Secii}) differ.
For the split graph of a bipartite lattice $X$, the flat bands always touch dispersive bands; in Fig.\,\ref{fig7}(a) we show the Lieb lattice $S(4)$ as an example.
In this lattice, the atoms highlighted by black dots (squares) belong to set $U$ ($V$). Within the unit cell, we have $1$ atom in $U$ and $2$ in $V$.

The general Hamiltonian of the split graph reads,
\begin{eqnarray}
H^\mathrm{split}\left(\boldsymbol{k}\right)	=	\left(\begin{array}{cc}
\mathbf{0}_{n^U\times n^U} & H\left(\boldsymbol{k}\right)_{n^U\times n^V}\\
H\left(\boldsymbol{k}\right)_{n^V\times n^U}^{\dagger} & \mathbf{0}_{n^V\times n^V}
\end{array}\right),\label{hamisplit}
\end{eqnarray}
where $n^U$ and $n^V$ are the numbers of vertices in $U$ and $V$, respectively.
Then, the rank of the Hamiltonian \ref{hamisplit} is $|n^U-n^V|$, indicating the existence of a set of flat bands at $E=0$ with degeneracy equal to $|n^U-n^V|$.
Here, taking the split graph of the square lattice, split graph of the honeycomb lattice as examples, we propose that the flat bands in split graphs of bipartite lattices can be topologically non-trivial when SOC is added.

We start from the flat band in $S(4)$. The lattices and corresponding band structure are shown in Fig.\,\ref{fig7} (a)-\textit{i} and (a)-\textit{ii}.
In the spinless case, there are three bands in total.
The lattice model of $S(4)$ can be built by putting $s$ orbitals at the $1a$ and $2e$ Wyckoff positions of space group $P4/mmm$.
The BR of the full set of bands is a sum of EBRs, given by $\left(A_{1g}\right)_{1a}\uparrow G\oplus\left(A_{g}\right)_{2e}\uparrow G$.
This means that the triple-degenerate point is not protected by the crystal symmetries.
One can add any symmetry-preserving perturbation, \textit{i.e.} on-site energy at the $1a$ Wyckoff position, to break the degeneracy and split the full set of bands into two branches, as shown by red lines in Fig.\,\ref{fig7} (a)-\textit{ii}.
The symmetry-data vectors of the two branches of bands read $\left(A_{1g}\right)_{1a}\uparrow G$ and $\left(A_{g}\right)_{2e}\uparrow G$, respectively.
The flat band and the upper dispersive band touch, with symmetry-data vector $\left\{\Gamma_{1}^{+}\oplus\Gamma_{2}^{+},\mathrm{M}_{5}^{-},\mathrm{X}_{1}^{+}\oplus\mathrm{X}_{4}^{-}\right\}$.
The parity of the flat band is $+1$ at $\Gamma$ and $-1$ at $\mathrm{M}$.
Thus, when the band touch is removed by SOC, the quasi-flat band cannot be topologically trivial.

In the spinful case, the symmetry-data vector of the top branch of bands is exactly a single EBR $\left(E_{g}\right)_{2e}\uparrow G$.
Then, if the inversion symmetry is maintained, any type of SOC that can break the band degeneracy will make the quasi-flat band topological.
Here, we add Rashba SOC that preserves the inversion symmetry, schematically shown in Fig.\,\ref{fig7}(a)-\textit{i}.
The bands are then gapped at the $\mathrm{M}$ point as shown in Fig.\,\ref{fig7}(a)-\textit{iii}.
The Wilson loop winding of the quasi-flat band shown in Fig.\,\ref{fig7}(a)-\textit{iv} indicates that the quasi-flat bands possess strong topology.

Now we turn to the $S(6)$ lattice. This lattice belongs to the space group $P6/mmm$, and the $2c$ and $3f$ positions are occupied as shown in Fig.\,\ref{fig7}(b)-\textit{i}.
Similar to what we found in the $S(4)$ lattice, the BR of the full set of bands is also a sum of EBRs: $\left(A_{1}'\right)_{2c}\uparrow G\oplus\left(A_{g}\right)_{3f}\uparrow G$.
With on-site energy added to the atoms located at $2c$, the band structure splits into two branches, see in Fig.\,\ref{fig7}(b)-\textit{ii}.
The symmetry-data vector of the upper branch bands is $\left\{\Gamma_{1}^{+}\oplus\Gamma_{5}^{+},\mathrm{K}_{1}+\mathrm{K}_{5},\mathrm{M}_{1}^{+}\oplus\mathrm{M}_{3}^{-}\oplus\mathrm{M}_{4}^{-}\right\}$.
The irrep of the quasi-flat band at the $\mathrm{M}$ point is $\mathrm{M}_{3}^{-}$.
The parities are $-1$ at $\mathrm{M}$ and $+1$ at $\Gamma$.
Thus, a topologically non-trivial phase is expectated upon adding SOC.
We introduce the inversion-symmetry-preserving Rashba SOC, and the resulting band structure with SOC is given by Fig.\,\ref{fig7}(b)-\textit{iii}.
The band touch will be removed, and the band topology of the quasi-flat is diagnosed by the odd Wilson loop winding in  Fig.\,\ref{fig7}(b)-\textit{iii}.

\section{Extensions of Our Work}\label{AppF}
The finding of topological flat bands in split graph of bipartite lattice inspire us to extend our work to all bipartite lattices that keep $|n^U-n^V|>0$. The dice lattice, shown in Fig.\,\ref{fig7}(c)-\textit{i}, is such kind of bipartite lattice.

As the band structure of the dice lattice shows (dashed blue lines in Fig.\,\ref{fig7}(c)-\textit{ii}), the system possesses one flat band that touches dispersive bands at the $\mathrm{K}$ point.
The symmetry-data vector of the full set of bands is $\left\{2\Gamma_{1}^{+}\oplus\Gamma_{4}^{-},\mathrm{K}_{1}\oplus\mathrm{K}_{5},2\mathrm{M}_{1}^{+}\oplus\mathrm{M}_{4}^{-}\right\}$ which is a sum of EBRs, \textit{i.e.} $\left(A_{1g}\right)_{1a}\uparrow G \oplus \left(A_{1}'\right)_{2c}\uparrow G$.
For the flat band, the irreps are $\Gamma_{4}^{-}$ at $\Gamma$ and $\mathrm{M}_{4}^{-}$ at $\mathrm{M}$.
The parities of the flat bands at all TRIMs are $-1$, which means that we cannot find any strong topology.
In fact, with an on-site energy perturbation, we can also split the full three bands into two branches, shown in red in Fig.\,\ref{fig7}(c)-\textit{ii}.
The symmetry-data vector of the upper two sets of bands is the single EBR $\left(A_{1}'\right)_{2c}\uparrow G$.
Thus, once the bands are gapped, the quasi-flat band cannot be topologically trivial.
The band structure of the dice lattice with Rashba SOC is shown in Fig.\,\ref{fig7}(c)-\textit{iii}.
The Rashba SOC breaks the inversion symmetry but keeps the $C_6$ symmetry.
With this kind of SOC added, the space group of this system will be reduced from $P6/mmm$ to $P6$.
The symmetry-data vector of the quasi-flat bands is $\left\{\bar{\Gamma}_{9}\oplus\bar{\Gamma}_{12},\mathrm{2K}_{4},\mathrm{M}_{3}\oplus\mathrm{M}_{4}\right\}$. This symmetry-data vector is not a sum but a difference of EBRs as $\left(^{1}\bar{E}^{2}\bar{E}\right)_{2b}\uparrow G\ominus\left(\right.$$^{1}$$\bar{E}_{3}$$^{2}$$\bar{E_{3}}$$\left.\right)_{1a}\uparrow G$ of the double space group $P6$.
Thus, the quasi-flat bands are a fragile topological state.
The non-trivial topological quasi-flat bands in the dice lattice can be also characterized by the Wilson loop.
As shown in Fig.\,\ref{fig7}(c)-\textit{iv}, there are four Wilson loop windings protected by the topology of the quasi-flat bands.

\end{appendix}

\end{document}